\documentclass[twocolumn]{jpsj3}
\usepackage{graphicx}
\usepackage{booktabs}
\usepackage{bm}
\newcommand{\nn}{\nonumber}

\def\sgn{\mathrm {sgn}}

\title{Singular and Half-Quantum Vortices and Associated Majorana Particles in Superfluid $^3$He-A between Parallel Plates}

\author{Takuto \textsc{Kawakami}\thanks{E-mail address: kawakami@mp.okayama-u.ac.jp}, Yasumasa \textsc{Tsutsumi}, and Kazushige \textsc{Machida} 
}
\inst{Department of physics, Okayama University, Okayama 700-8530 
}
\recdate{\today}
\abst{
	  Motivated by a recent experiment on superfluid $^3$He-A confined in narrow parallel plates using a rotating cryostat, 
	  we explore possible vortices stable under magnetic field applied to arbitrary angle relative to the plates
	  in order to seek vortices which can accommodate the Majorana zero mode in the core.
	  After proving that the singular vortex with the unit winding number provides 
	  the Majorana mode in the spinful situation, 
	  we establish the phase diagram in the plane; the rotation frequency $\Omega$ vs system size $R$ 
	  by finding possible order parameter textures within the Ginzburg-Landau framework.
	  We also analyze the stability for a single and a pair of half-quantum vortices, 
	  which possesses the Majorana mode in its core.
	  It is concluded from the above mention 
	  that the Majorana zero mode can be found in the present on-going experimental setting at ISSP, Univ. Tokyo.
	  }

\kword{$^3$He, A-phase, chiral $p$-wave pairing, superfluid, parallel plate geometry, vortex, texture}

\begin{document}
\maketitle

\section{Introduction}

There has been much attention on topological 
orders and the associated Fermionic quasi-particles with low energies~\cite{fu, ghaemi, nilsson, bargman}.
This is particularly true  when the quasi-particles are Majorana~\cite{majorana} character, namely
its creation operator $\gamma^{\dagger}$ is equal to its annihilation operator $\gamma$;				
$\gamma^{\dagger}=\gamma$. This unusual and intriguing Fermion, 
quite different from the usual Dirac particle, is thought to be
useful for fault-tolerant quantum computations
because it obeys non-Abelian statistics~\cite{ivanov}
and its existence is protected topologically to avoid decoherence.
These situations are ideal for quantum computation~\cite{dassarma}.
The candidate systems, which support the Majorana particle, are quite rare;
chiral spinless $p_x\pm ip_y$ superconductors, $p$-wave Feshbach resonanced superfluid~\cite{mizushima}, 
and the fractional quantum Hall state with the 5/2 filling.
The former superconductors have not been identified yet in nature.
 It has often been argued that Sr$_2$RuO$_4$
may be a prime candidate~\cite{maki,sarma,chung,vakaryuk}, but strong doubt has been cast on this possibility of Sr$_2$RuO$_4$ of 
its triplet pairing~\cite{machida2,lebed,mazin}.
Note that the first discovered triplet superconductor UPt$_3$ is an
$f$-wave pairing, not chiral $p$-wave~\cite{machida_a, machida_b, sauls_b}.
It is proven theoretically that the half-quantum vortex (HQV) in the chiral superconductors with
the $p$-wave pairing possesses the Majorana particle with zero energy localized at the vortex core~\cite{tewari, volovik}.

Superfluid $^3$He-A phase is characterized by a
chiral $p$-wave pairing.
There is no doubt on this identification~\cite{leggett, Vollhardt}. In fact, Volovik and Mineev~\cite{mineev}
are the first to point out the possibility to
the realization of HQVs in 1976. Since then, there have been
several general arguments on the stability of a HQV in connection with
$^3$He-A phase~\cite{cross, salomaa,salomaaRMP,volovikbook}. However, there are no
serious calculations which consider it in a realistic situation in
superfluid $^3$He-A phase on how to stabilize it and on what boundary conditions are needed for it.

Recently, Yamashita, $et$ $al.$~\cite{yamashita} have performed an experiment intended
to observe HQVs in superfluid $^3$He-A in parallel plate geometry.
The superfluid is confined in a cylindrical region with the radius $R=1.5$ mm
and the height 12.5 $\mu$m sandwiched by parallel plates.
A magnetic field $H=26.7$ mT ($\parallel$$\bm{z}$) is applied
perpendicular to the parallel plates
under pressure $P$=3.05 MPa.
Since the gap 12.5 $\mu$m
between plates is narrow compared to the dipole coherence length
$\xi_d\sim 10 \mu$m, the $l$-vector, which signifies the direction of orbital 
angular momentum of Cooper pairs,  is always perpendicular to the plates.
Also the $d$-vector is confined within the plane by an applied field $\bm{H} \parallel \bm{z} $
because the dipole magnetic field $H_d\sim 2.0 $ mT~\cite{leggett, Vollhardt}, where $\bm{H}$ tends to align the
$d$-vector perpendicular to the field direction.
They investigate to seek out various parameter spaces, such as
temperature $T$, or the rotation speed $\Omega$ up to $\Omega=6.28$ rad/sec
by using the rotating cryostat in ISSP, Univ. Tokyo, capable for the maximum rotation speed $\sim$12 rad/sec,
but there is no evidence for the existence of the HQV~\cite{yamashita}.

The aims in this paper are to investigate the possible vortex structures which can accommodate
the Majorana particle in the core under the above realistic experimental situations at ISSP 
for superfluid $^3$He-A phase confined in the parallel plates.
The candidate vortex structures in this situation are either the singular vortex (SV) with odd winging number 
or HQVs mentioned above.
Thus after examining the sufficient condition for the Majorana zero energy particle to exist in the
SV, we determine the phase diagram in the system size with the radius $R$ and the external rotation 
frequency $\Omega$ under applied fields with arbitrary angle relative to the plates.
Note that the external field is necessary for performing NMR detection~\cite{yamashita}.

The arrangement of the paper is as follows: In \textsection \ref{General} we examine the possible order parameters and its spatial structures, or textures.
By utilizing general symmetry properties of the superfluid $^3$He-A phase, we demonstrate that the SV with odd winding number even in the spinful situation can accommodate the Majorana zero mode in the core.
In \textsection \ref{SV} we investigate the stable vortices and textures within the Ginzburg-Landau (GL) formalism when the magnetic field is applied to an arbitrary angle relative to the plates, in order to find the stable SV region for various system parameters, external rotation $\Omega$, the system size $R$ and the field orientation $\theta _H$.
In \textsection \ref{HQV} the stability problem of the single HQV and a pair of HQVs are analyzed within the same framework.
We devote to summary and conclusions in the final section.

A part of the \textsection \ref{HQV} is published in ref. \ref{kawakamihqv}.

\section{General Considerations}\label{General}

\subsection{Order parameters and textures}
Generally, in $p$-wave superfluids, the order parameter (OP) is described by
\begin{eqnarray}
	\hat{\Delta}(\bm{r},\hat{\bm{p}})=
		\left(
		\begin{array}{cc}
			\Delta _{\uparrow   \uparrow}(\bm{r},\hat{\bm{p}}) & \Delta _{\uparrow	\downarrow} (\bm{r},\hat{\bm{p}}) \\
			\Delta _{\downarrow \uparrow}(\bm{r},\hat{\bm{p}}) & \Delta _{\downarrow \downarrow}(\bm{r},\hat{\bm{p}})
	    \end{array}
		\right).
\end{eqnarray}
The matrix element is described by 
\begin{eqnarray}
	\nn \Delta _{\sigma \sigma '}(\bm{r},\hat{\bm{p}}) &=&  A_{\sigma \sigma',+}(\bm{r})\hat{p}_+
												      + A_{\sigma \sigma',-}(\bm{r})\hat{p}_- \\ 
												   & &+ A_{\sigma \sigma',z}(\bm{r})\hat{p}_z,
\end{eqnarray}
where $\sigma$ is spin index $\uparrow$ or $\downarrow$ , $\hat{p} _{\pm}=\mp (\hat{p}_x \pm i\hat{p}_y)/\sqrt{2}$, 
and $\hat{\bm{p}}$ is the unit vector in the momentum space. 
In the parallel plate geometry, the momentum $\hat{p}_z$ component is suppressed. 
Here each component $A _{\sigma \sigma',\pm }(\bm{r})$ $(\sigma = \uparrow,\downarrow)$ can 
have its own phase winding whose winding number is denoted by $w _{\sigma \sigma', \pm}$. 
There are three possible textures at rest and under rotation as shown in Table \ref{winding}: $(w _{\sigma \sigma',+},w _{\sigma \sigma',-})=(0,2)$; A-phase texture (AT), (1,3); SV, and $(w _{\uparrow \uparrow,+},w _{\downarrow \downarrow,+},w _{\uparrow \uparrow,-},w _{\downarrow \downarrow,-})=(0,1,2,3)$; HQV. 
These textures and vortices are allowed in axisymmetric situation (see \textsection \ref{Hpara}).
We investigate the detailed configuration of these textures, the relative energetics, and the associated Majorana quasi-particle in the last two textures.
\begin{table}[tb]
\begin{tabular}{lcccccc} \hline\hline
 & $w _{\uparrow   \uparrow,   +}$ & $w _{\uparrow   \downarrow, +}$ &
   $w _{\downarrow \downarrow, +}$ & $w _{\uparrow   \uparrow,   -}$ &
   $w _{\uparrow   \downarrow, -}$ & $w _{\downarrow \downarrow, -}$ \\
\hline
AT  & 0 & 0 & 0 & 2 & 2 & 2 \\
SV  & 1 & 1 & 1 & 3 & 3 & 3 \\
HQV & 0 & $\times$ & 1 & 2 & $\times$ & 3 \\
\hline\hline
\end{tabular}
\caption{The combination of phase windings of each OP component for the A-phase texture (AT), the singular vortex (SV) and the half-quantum vortex (HQV). } \label{winding}
\end{table}
%
%
%
\subsection{Majorana bound state -symmetry considerations-} \label{Majorana}
\begin{figure}[tb]
	\begin{center}
		\includegraphics[width=60mm]{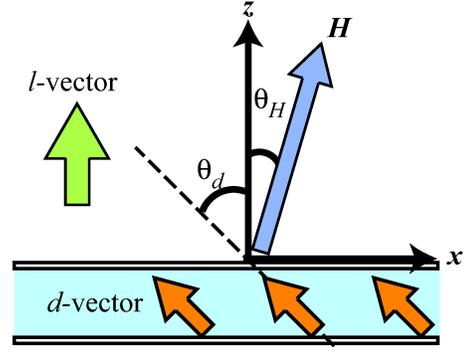}
		\caption{
				(Color online)
				Schematic diagram of the parallel plate system. 
				The external magnetic field $\bm{H}$ tilts by $\theta _H$ from the $z$-axis.
				$\theta_d$ is the polar angle of the $d$-vector.
				The external field tends to align the $d$-vector perpendicular to $\bm{H}$ 
				and the dipole interaction tends to aligned the $d$-vector parallel to the direction $\bm{z}$.
				Then $0\leq \theta_d\leq \pi/2-\theta _H$.
				}
		\label{def_theta1}
	\end{center}
\end{figure}

In this subsection, we examine the existence of the Majorana particle in the SV and the HQV in a most general situation. 
By assuming that the configurations of the OP are the bulk A-phase, the SV texture is described as 
\begin{eqnarray}
	 \nn \Delta _{\sigma \sigma '}(\bm{r},\hat{\bm{p}}) &=& A_{\sigma \sigma',+}(\bm{r})\hat{p}_+ \\
    	                                           &=& d_{\sigma \sigma'}(\bm{r})A_+(r)\exp(i\phi)\hat{p}_+,
\end{eqnarray}
where $\phi$ is the azimuthal angle of the center-of-mass coordinate $\bm{r}$,
\begin{eqnarray}
	\nn d_{\uparrow \uparrow(\downarrow \downarrow)}(\bm{r}) &=& \mp \frac{1}{\sqrt{2}}(d_x(\bm{r}) \mp id_y(\bm{r})) , \\
	\nn d_{\uparrow \downarrow }(\bm{r})                      &=& d_z(\bm{r}), 
\end{eqnarray} 
and $d$-vector $\bm{d}=(d_x,d_y,d_z)$ is real. 
Under this assumption, the direction of $d$-vector is determined by the dipole interaction and the interaction with the external field. 
The polar and azimuthal angle of direction of $d$-vector are determined as
\begin{eqnarray}
	\theta _d ^{(bulk)} &=& \frac{1}{2}\tan ^{-1} \left [\frac{\sin 2\theta _H}{(H_d/H)^2-\cos 2\theta _H}\right], 
	\label{bulkdt}\\
	\phi _d ^{(bulk)}  &=& 0. \label{bulkdp}
\end{eqnarray}
The angle $\theta _d$ and $\theta _H$ are defined as shown in Fig. \ref{def_theta1}.
The angle $\phi _d$ is the azimuthal angle of the direction of $d$-vector.
In other words, when the direction of the spin quantization axis is in the angle of
$(\theta _d ^{(bulk)}+\pi/2,\phi _d ^{(bulk)})$, $d_{\uparrow \downarrow }(\bm{r})=0$ throughout the system. 

When $\Delta _{\uparrow \downarrow }(\bm{r_1},\bm{r}_2)=\Delta _{\downarrow \uparrow }(\bm{r},\hat{\bm{p}})=0$, 
$p$-wave mean field Hamiltonian is written as
\begin{eqnarray}
	\mathcal{H}=\int d \bm{r}_1 d \bm{r}_2 \bm{\Psi }^\dagger(\bm{r_1})
	\left(
	\begin{array}{cc}
		\hat{\mathcal{K}} _{\uparrow \uparrow }  &0                                       \\
		0 	                                    & \hat{\mathcal{K}} _{\downarrow \downarrow }
	\end{array}
	\right)
	\bm{\Psi}(\bm{r_2}),
\end{eqnarray}
where 
\begin{eqnarray}
	\nn \bm{\Psi}(\bm{r})=[\psi _\uparrow  (\bm{r}), \psi _\uparrow  ^\dagger(\bm{r})
						  ,\psi _\downarrow(\bm{r}), \psi _\downarrow^\dagger(\bm{r})]^T, 					  
\end{eqnarray}
\begin{eqnarray}
	\nn \hat{\mathcal{K}}_{\sigma \sigma }=
	\left[
	\begin{array}{cc}
		H _0 ^\sigma (\bm{r}_1,\bm{r}_2)                 & \Delta _{\sigma \sigma}(\bm{r}_1,\bm{r}_2) \\
		-\Delta _{\sigma \sigma } ^*(\bm{r}_1,\bm{r}_2) & -H_0 ^{\sigma*} (\bm{r}_1,\bm{r}_2)          \\
	\end{array}
	\right], 
\end {eqnarray}
\begin{eqnarray}
\nn H_0^{(\sigma)}(\bm{r}_1,\bm{r}_2)&=&
	\left[
		-\frac{\nabla _1 ^2}{2m_3}+V(\bm{r}_1)-\mu_\sigma \right.\\
\nn & &  \left. +i\Omega \left(x_1 \frac{\partial }{\partial y_1}-y_1\frac{\partial }{\partial x_1} \right)
	\right]
	\delta(\bm{r}_1-\bm{r}_2),
\end{eqnarray}
where $m_3$, $V$, and $\mu _\sigma$ are the mass of the $^3$He atom, the single particle potential, 
and the chemical potential of the particle whose spin $\sigma$.
We carry out the Bogoliubov transformation to quasi particles whose creation and annihilation operators
\clearpage
\begin{eqnarray}
	\nn \eta _{\nu \uparrow } &=& 
	\int d \bm{r} 
	\left[ 
	      u_{\nu 1}^*(\bm{r})\psi _\uparrow           (\bm{r})
	    + u_{\nu 2}^*(\bm{r})\psi _\downarrow         (\bm{r}) 
	\right.\\ 
	& & \left.
		+ v_{\nu 1}^*(\bm{r})\psi _\uparrow   ^\dagger(\bm{r})
		+ v_{\nu 2}^*(\bm{r})\psi _\downarrow ^\dagger(\bm{r})
	\right], \\
	\nn \eta _{\nu \downarrow } &=& 
	\int d \bm{r} 
	\left[ 
	      v_{\nu 1}(\bm{r})\psi _\uparrow           (\bm{r})
	    + v_{\nu 2}(\bm{r})\psi _\downarrow         (\bm{r})
	\right.\\ 
	& & \left.
	    + u_{\nu 1}(\bm{r})\psi _\uparrow   ^\dagger(\bm{r})
		+ u_{\nu 2}(\bm{r})\psi _\downarrow ^\dagger(\bm{r})
	\right].
\end{eqnarray}
Obtained Bogoliubov-de Gennes (BdG) equation is
\begin{eqnarray} \label{BdG}
\nn \int d\bm{r}_2
	\left[
	\begin{array}{cc}
		\hat{{\mathcal K}}_{\uparrow \uparrow }(\bm{r}_1,\bm{r}_2)  &0  \\
	    0	                                                       & \hat{{\mathcal K}}_{\downarrow \downarrow }(\bm{r}_1,\bm{r}_2)
	\end{array}
	\right]
	\underline{u}_\nu(\bm{r_2})
	= \\
	\underline{u}_\nu (\bm{r}_1)
	\left[
	\begin{array}{cc}
		\hat{E}_\nu^\uparrow & 0 \\
    	0   	             & \hat{E}_\nu^\downarrow 
	\end{array}
	\right],
\end{eqnarray}
where $\hat{E}_\nu ^\sigma = \sigma _3 E_\nu^\sigma$.
This $\sigma _3$ is the Pauli matrix.
We can reduce the BdG equation (\ref{BdG}) to four eigenvalue equations,
\begin{eqnarray}
\nn \int d\bm{r}_2
	\hat{{\cal K}}_{\uparrow \uparrow }(\bm{r}_1,\bm{r}_2)
	\left(
	\begin{array}{c}
		u_{\nu 1}(\bm{r}_2) \\
		v_{\nu 1}(\bm{r}_2)
	\end{array} \right) 
	= \\
	\sgn(\sigma)E_\nu^{\sigma }
	\left(
	\begin{array}{c}
		u_{\nu 1}(\bm{r}_1) \\
		v_{\nu 1}(\bm{r}_1)
	\end{array} \right), 
\end{eqnarray}
\begin{eqnarray}
\nn \int d\bm{r}_2
	\hat{{\cal K}}_{\downarrow \downarrow }(\bm{r}_1,\bm{r}_2)
	\left(
	\begin{array}{c}
		u_{\nu 2}(\bm{r}_2) \\
		v_{\nu 2}(\bm{r}_2)
	\end{array} \right) 
	= \\
	\sgn(\sigma)E_\nu^{\sigma }
	\left(
	\begin{array}{c}
		u_{\nu 2}(\bm{r}_1) \\
		v_{\nu 2}(\bm{r}_1)
	\end{array} \right). 
\end{eqnarray}
These eigenvalue equations are solved numerically~\cite{tsutsumi}. 
If $\Delta_{\sigma \sigma }(\bm{r},\hat{\bm{p}})$ has odd phase winding,
we can obtain the zero energy state ($E ^\sigma _\nu=0$) and 
$\left(u_{\nu,i}(\bm{r}),v_{\nu,i}(\bm{r})\right)=\left(v_{\nu,i} ^*(\bm{r}),u_{\nu,i}^*(\bm{r})\right)$ 
so that $\eta _{\nu \sigma } = \eta _{\nu\sigma } ^{ \dagger }$. 
That is, the Bogoliubov quasi-particles exhibit the Majorana character.

In the SV texture all components of the OP have the odd winding number.
Thus the SV has the Majorana quasi-particle with the external field toward any direction.
In the HQV texture, either $\Delta _{\uparrow \uparrow } (\bm{r},\hat{\bm{p}})$ or $\Delta _{\downarrow \downarrow } (\bm{r},\hat{\bm{p}})$ has the odd phase winding.
The component that has no phase windings does not have the lower excitation.
That is, the HQV has the lower excitation that is induced by the odd phase winding and these excitations contain the Majorana quasi-particle.
Consequently, both the SV and the HQV have the Majorana quasi-particle at these vortex cores. 

However, in the above analysis, we do not consider the influence of the vortex core and the edge of system. 
In these situation, the configuration of the OP is different from the bulk A-phase due to variance of the OP, mixture of minor components, and so on. 
Furthermore, the Majorana quasi-particle exists at the vortex core and the edge of system. 
Then we have to clarify OP textures by more realistic and serious calculation using GL theory.

\section{Ginzburg-Landau Functional and Phase Diagram} \label{SV}

\subsection{Ginzburg-Landau functional}
The GL free-energy functional invariant under gauge transformation, spin, and orbital space rotations is well established ~\cite{leggett, Vollhardt, salomaaRMP, sauls_a, greywall, kita, fetter, thuneberg, wheatley} and given by a standard form
\begin{eqnarray} \label{GLfun}
	f_{total}=f_{grad}+f_{bulk}+f_{dipole}+f_{field},
\end{eqnarray}
\begin{eqnarray}
\nn f_{grad} = K\left[(\partial _{i}^*A_{\mu j}^*)(\partial _{i}A_{\mu j}) + (\partial _{i}^*A_{\mu j}^*)(\partial _{j}A_{\mu i})\right. \\
\nn + \left.(\partial _{i}^*A_{\mu i}^*)(\partial _{j}A_{\mu i})\right],  
\end{eqnarray}
\begin{eqnarray}
 \nn f_{bulk} & = & -\alpha _{\uparrow   \uparrow }   A_{\uparrow   \uparrow  ,i }^*A_{\uparrow   \uparrow  ,i}
	                -\alpha _{\uparrow   \downarrow } A_{\uparrow   \downarrow,i }^*A_{\uparrow   \downarrow,i} \\
\nn         & &     -\alpha _{\downarrow \downarrow } A_{\downarrow \downarrow,i }^*A_{\downarrow \downarrow,i}  \\
\nn         & &    + \beta _1 A_{\mu i}^*A_{\mu i}^*A_{\nu j}A_{\nu j} 
				   + \beta _2 A_{\mu i}^*A_{\nu j}^*A_{\mu i}A_{\nu j} \\
\nn 		& &    + \beta _3 A_{\mu i}^*A_{\nu i}^*A_{\mu j}A_{\nu j} 
			    + \beta _4 A_{\mu i}^*A_{\nu j}^*A_{\mu j}A_{\nu i} \\
\nn			& &   + \beta _5 A_{\mu i}^*A_{\mu j}^*A_{\nu i}A_{\nu j}, 
\end{eqnarray}
\begin{eqnarray}
	\nn f_{dipole}=g_d(A_{\mu \mu }^*A_{\nu \nu } + A_{\mu \nu }^*A_{\nu \mu } - \frac{2}{3} A_{\mu \nu }^*A_{\mu \nu }), 
\end{eqnarray}
\begin{eqnarray}
	\nn f_{field}=g_m H_\mu A_{\mu i}^*H_\nu A_{\nu i}, 
\end{eqnarray}
where $\mu,i=x,y,z$ and $\partial_i=\nabla_i-(2im_3/\hbar)(\bm\Omega \times \bm r)_i$ $(\bm \Omega \parallel \bm{z})$ with $m_3$ being the mass of the $^3$He atom. 
The coefficients of quadratic terms 
\begin{eqnarray}\label{alpha}
	\alpha _{\sigma \sigma'}=\alpha_0(1-T/T_c+\sgn(\sigma +\sigma ')\Delta T/T_c),
\end{eqnarray}
where $\alpha_0=N(0)/3$.
The Zeeman effect splits the transition temperature $T_{c \uparrow }$ of the $\left| \uparrow \uparrow \right>$ pair 
and  $T_{c \downarrow }$ of the $\left| \downarrow \downarrow \right>$ pair~\cite{ambegaokar}. 
The split width $\Delta T \equiv (T_{c \uparrow } - T_{c \downarrow })/2$ in eq. (\ref{alpha}).
However, we consider the region that $\Delta T \ll T_c-T$ so-called weak field region until \textsection\ref{HQV}.
The coefficient of fourth order terms $\beta _i$~\cite{GLparameters} are obtained in ref. \ref{sauls} appropriate for the experiment at $P=3.05$ MPa.
We use the gradient coupling constant with weak coupling limit $K=7\zeta(3)N(0)(\hbar v_F)^2/240(\pi k_B T_c)$. 
The $g_d$~\cite{thuneberg} and the $g_m$ are the coupling constant of the dipole interaction and the interaction with external field respectively,
\begin{eqnarray}
	g_d=\frac{\mu _0}{40}\left[ \gamma \hbar N(0) \ln \frac {1.1339 \times 0.45 T_F}{T_c}\right],
\end{eqnarray}
\begin{eqnarray}
	g_m=\frac{7\zeta(3)N(0)(\gamma \hbar)^2}{48[(1+F^a_0)\pi k_B T_c]}.
\end{eqnarray}
The transition temperature under no magnetic field $T_c$, the density of states $N(0)$, Fermi velocity $v_F$, the permeability of vacuum $\mu _0$, 
the gyromagnetic ratio $\gamma$, Landau parameter $F_0 ^a$, and Fermi temperature $T_F$ are given by experiments~\cite{wheatley, greywall}. 
In order to carry out realistic calculations taking account of the configuration of the vortex core and the edge of the system, 
our formulation includes strong coupling corrections in the bulk terms where the A-phase is stable over the B-phase.
Thus the gradient terms are within the weak coupling limit $(\rho _{sp}/\rho _s=1)$
because there is no established method to properly take into account the Fermi liquid correction for the OP form generalized beyond the A-phase
(see for standard method ref. \ref{vollhardt}).
We minimize the GL functional eq. (\ref{GLfun}) where the external field is applied to arbitrary direction relative to the plates and absolute value. 
First, we consider two limits where the direction of the external magnetic field is parallel and perpendicular to the plates. 

\subsection{Parallel field $H_{||}$} \label{Hpara}
In the situation where the external field applies parallel to the plates, both the dipole interaction and the external field suppresses the components of 
the OP $\Delta _{\uparrow \uparrow }(\bm{r})$ and $\Delta _{\downarrow \downarrow }(\bm{r})$. 
In the bulk A-phase system, the two components $A _{\uparrow \downarrow, \pm }(\bm{r})$ survive. 
Since the phase winding of the OP components increases the  kinetic energy of Cooper pairs, 
it is energetically favorable at rest that large component of the OP has small winding number. 
Furthermore, in the cylindrical symmetric system, $w_{\uparrow \downarrow,+}=w_{\uparrow \downarrow -}-2$ due to orbital coupling effect~\cite{kawakami}. 
Under low rotations the stable texture is the AT whose phase winding $(w_{\uparrow \downarrow,+},w_{\uparrow \downarrow,-})=(0,2)$ 
and the dominant component of the OP is $A_{\uparrow \downarrow,+}(\bm{r})$.

Under the anti-clock wise rotation the component that has the positive phase winding gains
the rotating energy through a term $-w_{\uparrow \downarrow \pm}|A_{\uparrow \downarrow \pm}|^2$ in the GL functional eq. (\ref{GLfun}).
As the rotating speed $\Omega$ is higher, the free energy of the SV whose dominant component of the OP has winding number $w=1$
becomes lower than that of the AT. 
According to the above analysis, the candidates of stable SVs are $(w_{\uparrow \downarrow,+},w_{\uparrow \downarrow,-})=(-1,1)$ and (1,3). 
The $(-1,1)$ texture has smaller winding number than the $(1,3)$ texture, 
from the kinetic energy view point the $(-1,1)$ texture seems to be more stable than the$(1,3)$. 
However, under rotation the minor component $A_{\uparrow \downarrow -}(\bm{r})$ of $(1,3)$ texture gains
more rotating energy than the minor component $A_{\uparrow \downarrow +}(\bm{r})$ of $(-1,1)$ texture. 
Under rotation the stable texture is determined by competition of those two factors. 
The results of our calculation show that when the rotating speed is large enough to stabilize the SV not AT, $(1,3)$ texture is always more stable than $(-1,1)$ texture. 

\begin{figure}[tb]
	\begin{center}
		\includegraphics[width=80mm]{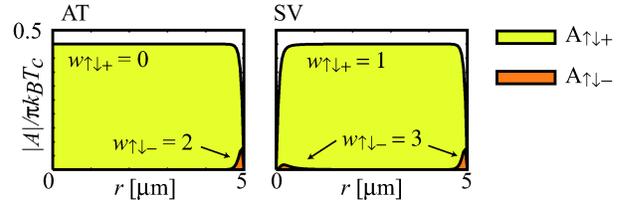}
		\caption{
				(Color online)
				Order parameter amplitude $\left| A_{\sigma \sigma', \pm }\right|$ normalized by $\pi k_B T_c$. 
				Left (right) figure shows the cross section of the OP components along the radial direction $\bm{r}$ of the AT (SV) texture 
				for $R=5$ $\mu$m, $T/T_c=0.95$, $H=10$ mT, $\theta _H=\pi/2$.
				We choose the spin quantization axis perpendicular to the plates; the $\bm{z}$ direction. 
				The spin component $\Delta _{\sigma \sigma }$ vanishes.
				}
		\label{tex_para}
	\end{center}
\end{figure}

In Fig. \ref{tex_para}, the cross section of the amplitude of the OP component of the AT and the SV 
when the direction of the external field is parallel to the plates and the quantization axis of spin is chosen perpendicular to the plates. 
Choosing the quantization axis parallel to the plates same as the direction of the field, 
we can obtain the SV texture whose OP component $\Delta _{\uparrow \downarrow }(\bm{r}) = 0$ exactly. 
The SV texture has Majorana quasi-particle state at the vortex core.
\begin{figure}[tb]
	\begin{center}
		\includegraphics[width=75mm]{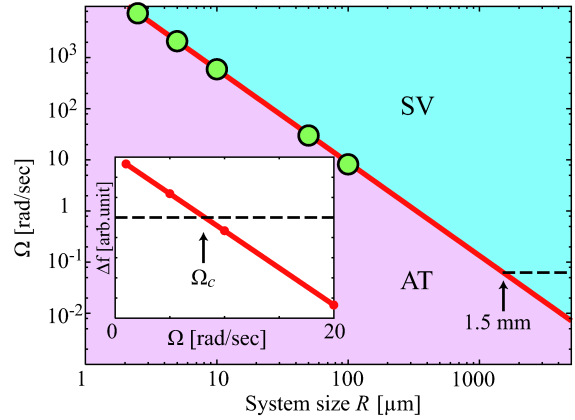}
		\caption{
				System size $R$ dependence of the critical angular velocity $\Omega _c$ from the AT to the SV 
				$T/T_c=0.95$, $\bm{H} \parallel \mathrm{plates}$ ($\theta _H =\pi /2$). 
				An extrapolated value of $\Omega _c = 0.06$ rad/sec is found at $R=1.5$ mm.
				The inset shows the energy differences $\Delta f$ between the two textures as a function of $\Omega$ for $R = 100$ $\mu$m.
				}
		\label{phase_diagram}
	\end{center}
\end{figure}

The phase diagram of textures in the space consisting of the rotating speed $\Omega$ and system size 
$R$ is shown in Fig. \ref{phase_diagram}  in a fixed temperature $T/T_c=0.95$. 
We calculate in various temperatures and absolute values of external field,
showing that the temperature and the absolute value of external field change $\Omega _c$ only slightly, so that $\Omega _c$ is determined by almost only $R$. 
We obtain the critical rotating speed $\Omega _c$ quantitatively as a function of system size $R$ smaller than 100 $\mu$m. 
Therefore, we can extrapolate $\Omega _c = 0.06$ rad/sec in $R = 1.5$ mm, that is the system size of the sample using the experiment in ISSP. 
This rotating speed can be well controlled by the present experimental technique.

\subsection{Perpendicular field $H_\perp$}
\begin{figure}[tb]
	\begin{center}
		\includegraphics[width=80mm]{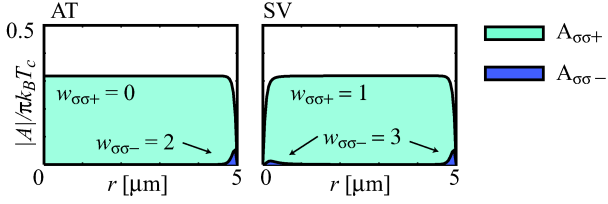}
		\caption{
				(Color online)
				Order parameter amplitude $\left| A_{\sigma \sigma', \pm }\right|$ normalized by $\pi k_B T_c$ 
				for $R=5$ $\mu$m, $T/T_c=0.95$, $H=10$ mT, $\theta _H=0$. 
				Left (right) figure shows the cross section of the OP components along the radial direction $\bm{r}$ of the AT (SV) texture.
				We choose the spin quantization axis perpendicular to the plates; $\bm{z}$ direction. 
				The spin component $\Delta _{\uparrow \downarrow }(\bm{r},\hat{\bm{p}})=0$ exactly. 
				}
		\label{tex_perp1}
	\end{center}
\end{figure}

In this subsection we consider the situation where the external field apply parallel to the plates.
When the external field larger than the dipole field $H_d$, the $d$-vectors lie parallel to the plates. 
The result of numerical calculation shows that $\Delta _{\uparrow \downarrow }(\bm{r},\hat{\bm{p}})=0$ 
exactly throughout the system as shown in Fig. \ref{tex_perp1}. 
These SV textures have the Majorana quasi-particle state at the vortex core. 
When the external field is smaller than the dipole field $H_d$, in the bulk A-phase the $d$-vector aligns perpendicular to the plates. 
The OP texture is same as in the case that the external field is parallel to the plates shown in Fig. \ref{tex_para}. 
Choosing the quantization axis parallel to the plates, we can obtain the OP component $\Delta_{\uparrow \downarrow }(\bm{r})=0$ exactly. 
However, when $d\mbox{-vector} \parallel \bm{H}$ taking account of Zeeman effect in BdG equation (\ref{BdG}), the symmetry of the equation arrows the solution $\left(u_{\nu,1}(\bm{r}),v_{\nu,1}(\bm{r})\right)=\left(v_{\nu,2} ^*(\bm{r}),u_{\nu,2}^*(\bm{r})\right)$.
Thus the core bound state of the SV cannot satisfy Majorana condition 
$\left(u_{\nu,i}(\bm{r}),v_{\nu,i}(\bm{r})\right)=\left(v_{\nu,i} ^*(\bm{r}),u_{\nu,i}^*(\bm{r})\right)$.
Therefore only in the case where the external field is larger than the dipole field $H_d$, the stabilized SV has the Majorana quasi-particle.

The phase diagram of textures are same as that in \textsection \ref{Hpara} since in the result of our calculation, the external magnetic field change critical rotating speed $\Omega _c$ of the texture transition from the AT to the SV only slightly  as shown in \textsection \ref{Harbitral}.

In the case that $\bm{H}\perp \mathrm{plates}$ where the external field is larger than the dipole field $H_d$, the $d$-vector is perpendicular to the $l$-vector. 
Then the dipole energy of the bulk system can be neglected.
Thus this case is favorable for realization of the HQV texture. 
We discuss the HQV texture in \textsection \ref{HQV} in detail.
%
\subsection{Arbitrary oriented field} \label{Harbitral}

\begin{figure}[tb]
	\begin{center}
		\includegraphics[width=80mm]{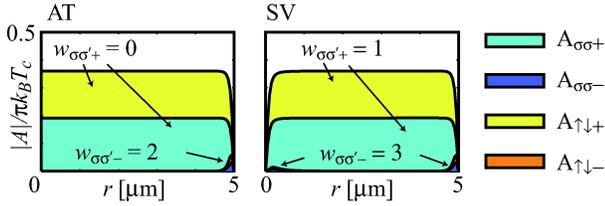}
		\caption{
				(Color online)
				Order parameter amplitude $\left| A_{\sigma \sigma', \pm }\right|$ normalized by $\pi k_B T_c$ 
				for $R=5$ $\mu$m, $T/T_c=0.95$, $H = 2$ mT, $\theta _H = \pi/18$. 
				Left (right) figure shows the cross section of the OP components along the radial direction $r$ of the AT (SV) texture.
				We choose the spin quantization axis perpendicular to the plates; $\bm{z}$ direction. 
				All spin components $\Delta_{\sigma \sigma'}(\bm{r},\hat{\bm{p}})$ are non-vanishing. 
				When we choose the spin quantization axis to be the direction $\theta=\theta_d^{(bulk)}+\pi/2$ and $\phi=0$,
				$A_{\uparrow \downarrow, \pm }\ll A_{\sigma \sigma, \pm }$.
				}
		\label{tex_10}
	\end{center}
\end{figure}

In the bulk A-phase system, the direction of the $d$-vector is determined by the competition between the dipole interaction 
favoring that the $d$-vector is parallel to $l$-vector and the external field favoring that the $d$-vector is perpendicular to the $\bm{H}$. 
The direction of the $d$-vector is given by eqs. (\ref{bulkdt}) and (\ref{bulkdp}). 
When the quantization axis is perpendicular to the plates, all components
$\Delta _{\sigma \sigma'}(\bm{r},\hat{\bm{p}})$ are non-vanishing as shown in Fig. \ref{tex_10}. 

\begin{figure}[tb]
	\begin{center}
		\includegraphics[width=80mm]{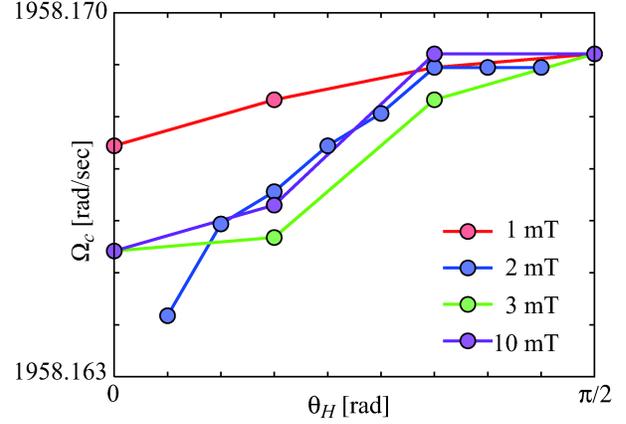}
		\caption{
				(Color online)
				The critical angular velocity $\Omega _c$ dependence of $\theta _H$
				from the AT to the SV for $T/T_c=0.95$, $R = 5$ $\mu$m, $H= 1$, $2$, $3$, and $10$ mT.
				Under the various direction and absolute value of the external fields, 
				$\Omega _c$ coincides with each other for the precision in second order after the decimal point.
				}
		\label{omega_th}
	\end{center}
\end{figure}

We consider the energetics of these textures by calculating the stable textures under
various external magnetic fields; $H$ = 1, 2, 3, and 10 mT, for various polar angles $\theta _H=n\pi/18$ $n=1,2, \cdots$. 
The result of the calculation in the system $R=5$ $\mu$m which is shown in Fig. \ref{omega_th}. 
Under these external fields, $\Omega _c$ coincides with each other for the precision in second order after the decimal point.
We see that the external magnetic field scarcely changes the phase diagram shown in Fig. \ref{phase_diagram}. 
Their reasons are as follows. 
The critical rotation $\Omega _c$ is determined by the competition of the kinetic energy and the energy gain from angular momentum, namely, the gradient term.  On the other hands the influence of the interaction with the external field is no more than $10^{-6}$ order of the gradient energy.

\begin{figure*}[tb]
	\begin{center}
		\includegraphics[width=160mm]{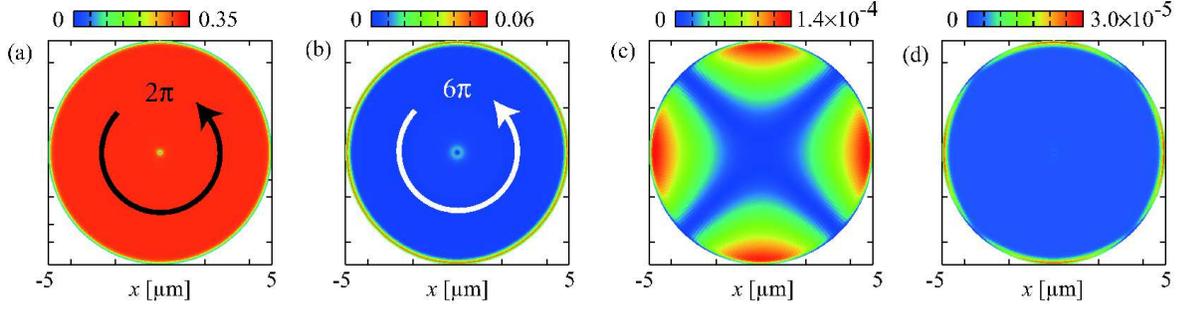}
		\caption{
				(Color online)
				Order parameter amplitude 
				(a) $\left| A_{\sigma   \sigma,     + }\right|$, 
				(b) $\left| A_{\sigma   \sigma,     - }\right|$, 
				(c) $\left| A_{\uparrow \downarrow, + }\right|$, and
				(d) $\left| A_{\uparrow \downarrow, - }\right|$ 
				normalized by $\pi k_B T_c$ for $R=5$ $\mu$m, $T/T_c=0.95$, $H = 2$ mT, $\theta _H = \pi/18$.
				We choose the spin quantization axis $\theta = \theta_d ^{(bulk)}+\pi/2$, $\phi = 0$. 
				The spin components $\Delta_{\uparrow \downarrow }(\bm{r},\hat{\bm{p}}) \neq 0$ but have the amplitude 
				with the order $10^{-3}$ of the amplitude of $\Delta_{\sigma \sigma }(\bm{r},\hat{\bm{p}})$ 
				and modulate spatially. 
				}
		\label{amp_minor}
	\end{center}
\end{figure*}

We discuss the possibility that the Majorana quasi-particle exists in these textures. 
As shown in \textsection \ref{Majorana}, the SV has the Majorana quasi-particle when their OP component 
$\Delta _{\uparrow \downarrow}(\bm{r},\hat{\bm{p}})=0$ for  appropriate quantization axis throughout the system. 
Thus we turn the quantization axis to be their polar and azimuthal angle $(0,0) \mapsto (\theta _d ^{(bulk)}+\pi/2, \phi _d^{(bulk)})$ in order to change the expression to that suppressing the component of the OP $\Delta _{\uparrow \downarrow}(\bm{r},\hat{\bm{p}})$. 
Then we obtain changed OP textures as shown in Fig. \ref{amp_minor}. 
In these textures, the components of the OP $A_{\uparrow \downarrow,\pm}(\bm{r})\neq 0$, 
but finite amplitude with the order $10^{-3}$ against the dominant component $A_{\sigma \sigma \pm}(\bm{r})$. 
These amplitudes vary spatially, and their phase cannot be specified by the accurate A-phase OP. 
In addition, when the $d$-vector is not perpendicular to the magnetic field $\bm{H}$, we should note the influence of Zeeman effect for the Majorana condition.
Therefore, we cannot conclude whether or not those SVs have the Majorana quasi-particle. 
In order to clarify these points, we have to solve spinful BdG equation include off-diagonal terms of eq. (\ref{BdG}). 
These remain as a future problem. 

\begin{figure}[tb]
	\begin{center}
		\includegraphics[width=75mm]{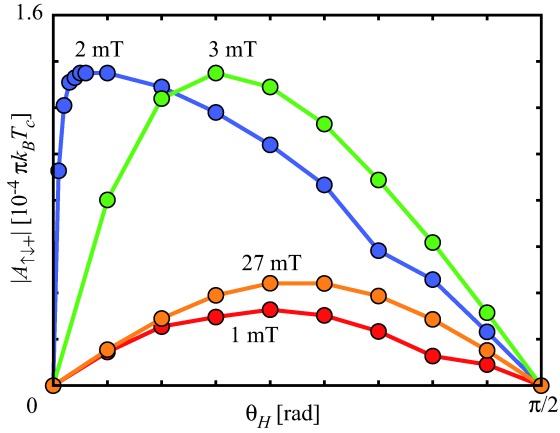}
		\caption{
				(Color online)
				The maximum value of $\left| A_{\uparrow \downarrow +} \right|$ as a function of $\theta_H$ 
				for $H = 2$, $3$, and $27$ mT, $R = 5$ $\mu$m, $T/T_c=0.95$. 
				When $\theta _H =0$ or $\pi/2$, 
				the amplitude of the component $\Delta_{\uparrow \downarrow }(\bm{r},\hat{\bm{p}}) =0$ exactly
				When the external field is comparable to dipole field ($\simeq 2$) mT, 
				the amplitude of these components is enhanced.
				}
		\label{amp_H}
	\end{center}
\end{figure}

We notice these induced components. 
In Fig. \ref{amp_H}, we show the maximum value of amplitude of induced component $A_{\uparrow \downarrow,+}(\bm{r})$ as a function of $\theta _H$ in various 
external field values. 
We show that when the direction of the external field are perpendicular to the plates ($\theta _H=0$) and parallel to the plates ($\theta _H=\pi/2$), the component $A_{\uparrow \downarrow,+}(\bm{r}) = 0$ exactly. 
Moreover, when the external field comparable to the dipole field $H_d$ ($\simeq 2$ mT), 
the induced component is enhanced maximally under such a magnitude of the field.

\begin{figure}[tb]
	\begin{center}
		\includegraphics[width=75mm]{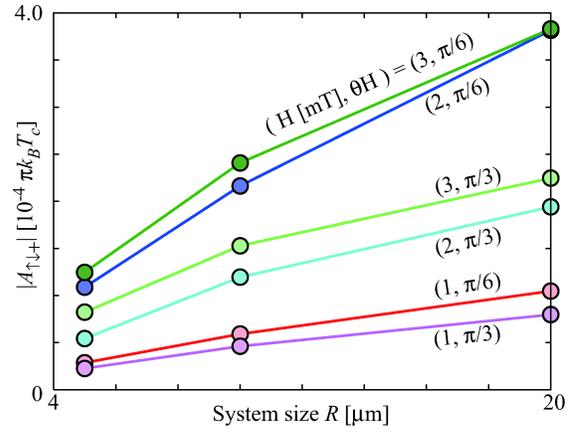}
		\caption{
				The maximum value of $\left| A_{\uparrow \downarrow +} \right|$ as a function of $R$ 
				for $H = 1$, $2$, and $3$ mT, $\theta _H = \pi/6$ and $\pi/3$ rad, $\mu$m, $T/T_c=0.95$. 
				When the system size becomes larger, the amplitude of these components becomes larger. 
				In the system using experiment, $R = 1.5$ mm, $\left| A_{\uparrow \downarrow \pm} \right|$ are non-vanishing.
				}
		\label{amp_size}
	\end{center}
\end{figure}

The system size dependence of them is shown in Fig. \ref{amp_size}. 
When the system size becomes large, the amplitude of the component $A_{\uparrow \downarrow,+}(\bm{r})$ tend to be large. 
Therefore, even if the large sample such as that used in experiment $R=1.5$ mm, these components are finite.

We suggest that the reason why the induced component is non-zero is due to the existence of induced $A_{\sigma \sigma',-}$ component. 
Taking account of the $A_{\sigma \sigma',-}$ component, the direction of $d$-vector determined by minimizing $f_{dipole} + f_{field}$ has the component 
parallel to the spin quantization axis except for the case that $\bm{H}\parallel$ plates and $\bm{H}\perp$ plates.
In addition, these components varies spatially according to spatial modulation of $A_{\sigma \sigma',-}$ component.

\section{Half-Quantum Vortex} \label{HQV}

\subsection{Single half-quantum vortex}
\begin{figure}[tb]
	\begin{center}
		\includegraphics[width=80mm]{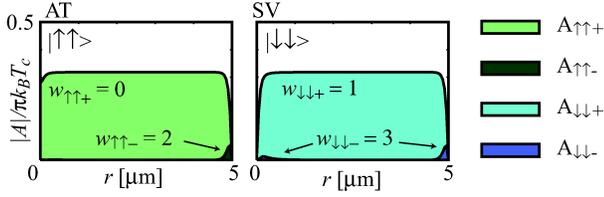}
		\caption{
				(Color online)
				Order parameter amplitude of
				(a) $A_{\uparrow   \uparrow,   \pm }$ and 
				(b) $A_{\downarrow \downarrow, \pm }$ components normalized by $\pi k_B T_c$ 
				for $R=5$ $\mu$m, $T/T_c=0.95$, $H = 10$ mT, $\theta _H = 0$.
				We choose the spin quantization axis perpendicular to the plates; $\bm{z}$ direction.
				At the vortex core, $A_{\uparrow \uparrow, \pm }$ component is only non-vanishing, 
				so that the HQV is A$_1$-core vortex.
				}
		\label{tex_HQV}
	\end{center}
\end{figure}

\begin{figure}[tb]
	\begin{center}
		\includegraphics[width=80mm]{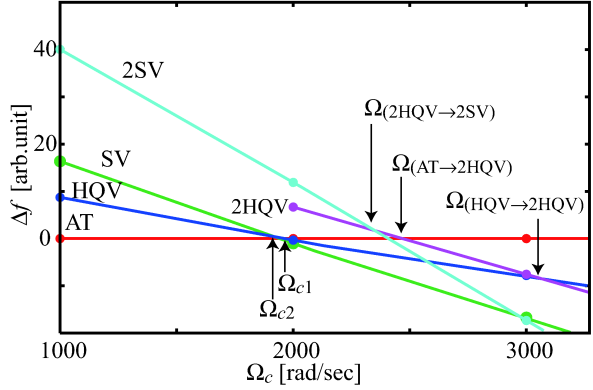}
		\caption{
				(Color online)
				Free-energy comparison for the AT, the HQV, the SV, the two HQVs system (2HQV), and the two SVs system (2SV) 
				as a function of $\Omega$ for $R = 10$ $\mu$m and $T/T_c=0.97$ 
				at the weak field region.
				The $\Delta f$ is the relative free energy to the AT.
				In the weak field region, $\Omega _{c2} < \Omega _{c1}$.
				}
		\label{HQV_df_omg}
	\end{center}
\end{figure}

When the direction of the external field is perpendicular to the plates, it is a favorable situation for the HQV to realize. 
Then in this section, we consider possible stable HQV texture and their energetics. 
First, we consider the case that there is a single HQV in the system. 
The external magnetic field splits $T_c$ of the $\left| \uparrow \uparrow \right>$ and $\left| \downarrow \downarrow \right>$ pairs by the Zeeman effect~\cite{ambegaokar}.
The splitting is defined as $\Delta T = (T_{c\uparrow }-T_{c\downarrow })/2$.
When $\Delta T > 0$ ($\Delta T < 0$), the amplitude of the OP $\Delta _{\uparrow \uparrow }$ ($\Delta _{\downarrow \downarrow }$) becomes larger than $\Delta _{\downarrow \downarrow }$ ($\Delta _{\uparrow \uparrow }$).
The splitting $\Delta T$ is generally much smaller than the amplitude of the OP in the so-called weak field region $H\simeq$ 10 mT. 
In this region, the texture of the HQV are shown in Fig. \ref{tex_HQV}. 
The rotation speed $\Omega _{c1}$ ($\Omega _{c2}$) at which the free energy of the HQV (SV) and the AT (HQV) are intersect. 
In Fig. \ref{HQV_df_omg}, we show that the HQV never becomes the absolute stable texture. 
The reasons are the strong coupling effect in the bulk fourth terms of GL functional eq. (\ref{GLfun}). 
In the weak coupling limit, under some critical rotation, the AT, the SV, and the HQV degenerate each other. 
The strong coupling effect stabilize the A phase, then the structure of the central region of the AT are more favorable than the others. 
Therefore, the critical rotation $\Omega _{c1}$ are larger than $\Omega _{c2}$. 

\begin{figure}[tb]
	\begin{center}
		\includegraphics[width=80mm]{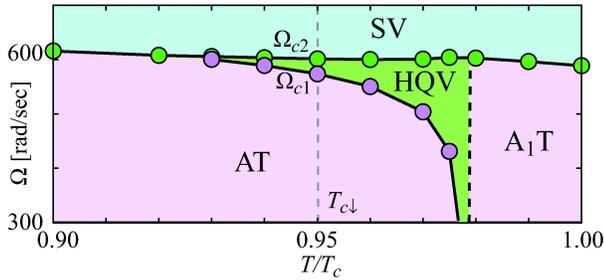}
		\caption{
				(Color online)
				Stability region of the HQV in $\Omega$ versus $T/T_c$ ($R = 10$ $\mu$m and $\Delta T/T_c = 0.05$). 
				A$_1$T denotes the A$_1$ phase texture where only $\left| \uparrow \uparrow \right>$ pair exist.
				The stability region of the HQV becomes wide at the vicinity of the transition temperature 
				from the A-phase to the A$_1$-phase.
				}
		\label{HQV_omg_t}
	\end{center}
\end{figure}

\begin{figure}[tb]
	\begin{center}
		\includegraphics[width=80mm]{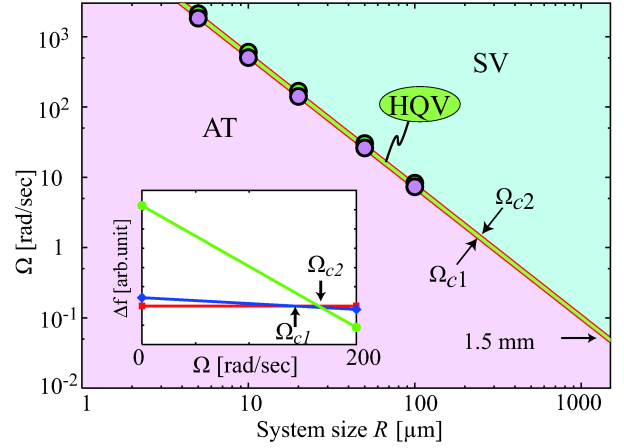}
		\caption{Stability region of HQV sandwiched between $\Omega _{c1}$ and $\Omega _{c2}$ 
		as a function of $R$ for $T/T_c = 0.97$ and $\Delta T/T_c = 0.05$ ($H\simeq 100$ mT). 
		The critical rotation $\Omega _{c1} = 0.05$ rad/sec is the extrapolated value for $R = 1.5$ mm.
		Inset shows the free-energy comparison for $R = 20$ $\mu$m, displaying the successive transitions 
		from the AT to the HQV at $\Omega _{c1}$ and from the HQV to the SV at $\Omega _{c2}$.
		}
		\label{HQV_omg_r}
	\end{center}
\end{figure}

When the temperature becomes larger, the amplitude of the OP becomes smaller.
Then the split $\Delta T$ cannot be neglected. 
When the split $\Delta T>0$, the $\left| \uparrow \uparrow \right>$ pair becomes dominant. 
Thus, the HQV texture and the energy difference $\Delta f$ in Fig. \ref{HQV_df_omg} become close to the AT. 
In this process, $\Omega _{c1}$ become smaller than $\Omega _{c2}$. 
Furthermore, when the temperature becomes larger, the stable region of the HQV becomes larger as shown in Fig. \ref{HQV_omg_t}. 
Consequently, the phase diagram is shown in Fig. \ref{HQV_omg_r}. 
We can extrapolate that the stability region of the HQV is $0.05 < \Omega < 0.06$ rad/sec in the sample whose size $R = 1.5$ mm.
We also notice that by changing the radius $R$ of the system one can control the width of the stability region.
For example, in $R=100$ $\mu$m, the stability region of the HQV is $7 < \Omega < 8$ rad/sec.

\subsection{A pair of half-quantum vortices}
We examine the case that there are more than two half-quantum vortices in the system. 
There are two possibilities to enter the two HQVs in the system. 
Namely, the one case is that there are the HQVs whose winding number $(w_{\uparrow \uparrow +}, w_{\downarrow \downarrow +}, w_{\uparrow \uparrow -}, w_{\downarrow \downarrow -}) = (0,1,0,3)$ and $(1,0,3,0)$ in the system, the other case is that there are two (0,1,2,3)-HQVs. 
We calculate these two cases and conclude that the former is not solution of our calculation. 
The center of phase winding of the component of the OP does not coincide with the center of the system, 
which is not the energetically advantageous form for the angular momentum.
Furthermore, there is no repulsion between HQVs because we assume that the gradient energy are the weak coupling limit form.

\begin{figure}[tb]
	\begin{center}
		\includegraphics[width=80mm]{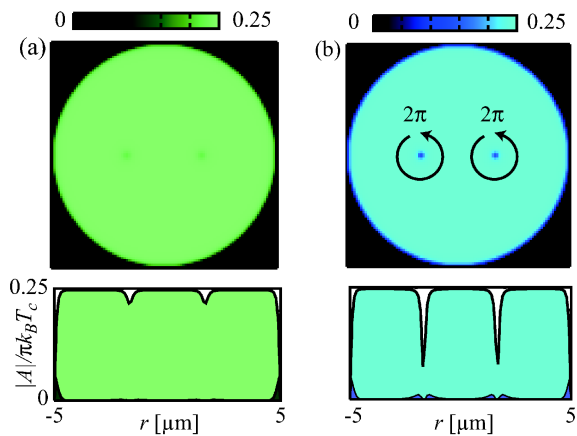}
		\caption{
				(Color online)
				Order parameter amplitude 
				(a) $A_{\uparrow   \uparrow  , \pm }$ and 
				(b) $A_{\downarrow \downarrow, \pm }$ components normalized by $\pi k_B T_c$ 
				for the metastable 2HQV system, $R=5$ $\mu$m, $T/T_c=0.95$, $\theta _H = 0$.
				We choose the spin quantization axis perpendicular to the plates; $\bm{z}$ direction.
				There are two vortices whose winding number 1 in the $\left| \downarrow \downarrow \right>$ space. 
				The sense of the phase winding and its number are shown.
				The low figures are cross sections of each component across the vortex cores.
				}
		\label{2HQV_amp}
	\end{center}
\end{figure}

We investigate the case that there are two (0,1,2,3)-HQVs in the system. 
In the Fig. \ref{2HQV_amp} we show the texture of the two HQVs system (2HQV). 
In this texture, the component that has the phase winding are the same spin state. 
Thus the distance between them is determined by the competition of the kinetic energy loss and the energy gain from the angular momentum. 
However, we conclude these are just metastable texture from our calculation. 

First of all, we consider the case that $\Delta T<0$ where the $\left| \downarrow \downarrow \right>$ pair is dominant. 
In this case, the texture of the 2HQV is close to two SVs system (2SV). 
From the analogy with the single HQV case, it can be guessed that critical rotation $\Omega _{(\mathrm{AT} \rightarrow \mathrm{2HQV})}$ 
where the free energy of the 2HQV cross to that of the AT is larger than the critical rotation $\Omega _{(\mathrm{2HQV} \rightarrow \mathrm{2SV})}$ 
where the free energy of the 2SV crosses to that of the AT. 
Namely, the 2HQV cannot be stable in this course. 

\begin{figure}[tb]
	\begin{center}
		\includegraphics[width=80mm]{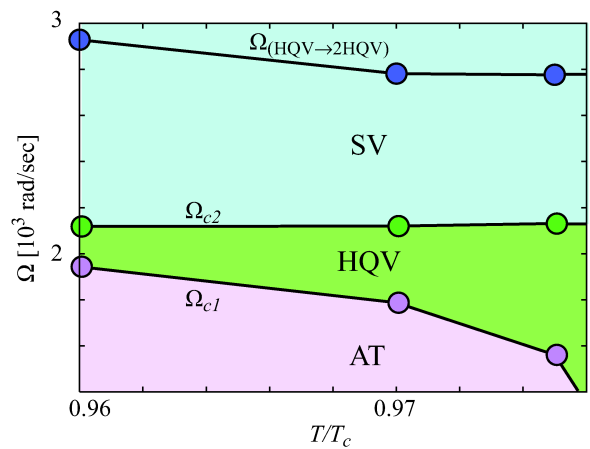}
		\caption{
				Critical angular velocity $\Omega _{(\mathrm{HQV} \rightarrow \mathrm{2HQV})}$ from the HQV 
				to the 2HQV for $R=5$ $\mu$m and $\Delta T/T_{c}$=0.05.
				Even if the temperature becomes higher, $\Omega _{(\mathrm{HQV} \rightarrow \mathrm{2HQV})}$ does not become lower than $\Omega _{c2}$.
				}
	    \label{2HQV_omg_t}
	\end{center}
\end{figure}

On the other hands, we consider the case that $\Delta T > 0$ where the $\left| \uparrow \uparrow \right>$ pair is dominant.
In this case the texture of the 2HQV are energetically close to the AT. 
From the analogy with the single HQV case, it is possible that $\Omega _{(\mathrm{AT} \rightarrow \mathrm{2HQV})}<\Omega _{(\mathrm{2HQV} \rightarrow \mathrm{2SV})}$. 
However, in these regions, the SV is more stable than the AT. From our calculation, we show that 
the critical rotation speed $\Omega _{(\mathrm{HQV} \rightarrow \mathrm{2HQV})}$ where the free energy of the 2HQV crosses to that 
of the HQV cannot smaller than the critical rotation where the free energy of the SV crosses to that 
of the HQV as shown in Fig. \ref{2HQV_omg_t}. Namely the 2HQV cannot also be stable in this course. 

\section{Summary an conclusions}
We have studied the relative stability of the three kinds of the order parameter textures; the A-phase texture, the singular vortex, and the half-quantum vortex for superfluid $^3$He-A phase confined by narrow parallel plates with the radius $R$.
After showing in \textsection \ref{General} that the spinful singular vortex can accommodate the Majorana quasi-particle localized at the core, we examine the condition that the Majorana quasi-particle exists, namely, the stability of the singular vortex and also half-quantum vortex which is known to accommodate the Majorana quasi-particle.

We have found that under the external field applied to either exactly parallel or perpendicular to the plates, the singular vortex state becomes stable above the critical rotation speed $\Omega _c$ which is evaluated. 
Thus the Majorana quasi-particles can be observed in those situation. The on-going experiment using the rotating cryostat at ISSP, Univ. Tokyo is particularly suited for this task because $\Omega _c \simeq 0.06$ rad/sec is within well controlled rotation speed. 
Moreover since singular vortex carries the Majorana quasi-particle localized at the core, we can design the braiding experiment by manipulating two or more Majorana quasi-particles.

Our calculations on the relative stability of the relevant textures are based on the standard Ginzburg-Landau functional which is firmly established through the cross checking between experiments and theories over several decades. Thus we believe that our results should be reliable not only qualitatively, but also quantitatively.

There are a few question remained to be clarified; (1) When the field is applied to exactly neither $\theta _H \neq 0$, or $\theta _H \neq \pi /2$, there appears the unwanted OP component $\Delta _{\uparrow \downarrow }$ whose magnitude is an order of $10^{-3}$ compared with the main component. 
At present we do not know whether or not this component is really harmful for the existence of the Majorana quasi-particle for the singular vortex.
(2) We do not know how to take into account the so-called strong coupling effect in the gradient term in eq. (\ref{GLfun}) for general order parameter state. This remains a future problem.

In conclusion, independent of the above two reservations (1) and (2) we emphasize the fact that the parallel plate geometry whose gap is an order of 10 $\mu$m for superfluid $^3$He-A phase is one of the best systems for detecting the Majorana quasi-particle.

\section*{Acknowledgments}
 We thank T. Mizushima and M. Ichioka for useful discussions.


\begin{thebibliography}{99}
\bibitem{GLparameters} The following GL parameters are
used: $\alpha_0=3.81\times10^{50}[J^{-1}m^{-3}]$, and $\beta_1=-3.75, 
\beta_2=6.65,\beta_3=6.56,\beta_4=5.99, \beta_5=-8.53$ in units of $10^{99}[J^{-3}m^{-3}].$
\end{thebibliography}
\end{document}